# An Intelligent Infrastructure as a Foundation for Modern Science

Satrajit S. Ghosh

McGovern Institute for Brain Research, Massachusetts Institute of Technology, Cambridge, MA

Department of Otolaryngology - Head and Neck Surgery, Harvard Medical School, Boston, MA

Correspondence: satra@mit.edu

## Abstract

Infrastructure shapes societies and scientific discovery. Traditional scientific infrastructure, often static and fragmented, leads to issues like data silos, lack of interoperability and reproducibility, and unsustainable short-lived solutions. Our current technical inability and social reticence to connect and coordinate scientific research and engineering lead to inefficiencies and impede progress. With AI technologies changing how we interact with the world around us, there is an opportunity to transform scientific processes.

Neuroscience's exponential growth of multimodal and multiscale data, together with its urgent clinical relevance, demands an adaptive infrastructure that can expose computable states, coordinate across systems, and improve through use. Using neuroscience as a stress test, this perspective argues for a paradigm shift: infrastructure must evolve into a dynamic, AI-aligned ecosystem to accelerate science. Building on several existing principles for data, collective benefit, and digital repositories, I recommend operational guidelines for implementing these principles to create this dynamic ecosystem, aiming to foster a decentralized, self-learning, and self-correcting system where humans and AI can collaborate seamlessly. Addressing the chronic underfunding of scientific infrastructure, acknowledging diverse contributions beyond publications, and coordinating global efforts are critical for this transformation. The capability that remains missing is not another repository, standard, or model, but a machine-actionable coordination layer that keeps independently governed systems continuously compatible as science changes. This coordinating role, even more than analysis, is where AI becomes transformative rather than merely assistive.

By prioritizing an intelligent infrastructure as a central scientific instrument for knowledge generation, we can overcome current limitations, accelerate discovery, ensure reproducibility and ethical practices, and ultimately translate neuroscientific understanding into tangible societal benefits, setting a blueprint for other scientific domains.

Infrastructure transforms societies[1]. It powers cities, connects communities, and fuels the engines of progress. From power grids and highways to satellites and the internet, it moves people, goods, information, and ideas. It also serves as the backbone of scientific advancement[2,3]. Particle accelerators, genome sequencers, and AI systems are examples of infrastructure that open new frontiers of knowledge, allowing us to peer into the origins of the universe, decode the building blocks of life, and transform our everyday lives. But as environmental, energy, and health challenges grow more complex and urgent, we will need to transform the world's science and engineering infrastructure to address them.

Scientific infrastructure refers to technical and sociotechnical systems that enable research communities to create, exchange, compute on, interpret, govern, and preserve scientific information over time. In neuroscience this includes data repositories, compute and cloud resources, standards and knowledgebases, workflow and analysis platforms, programmatic interfaces, governance mechanisms, and the human expertise needed to operate them. The term relates to persistence: ad hoc scripts, short-lived lab tools, and isolated websites are not themselves infrastructure unless they mature into durable, governed, and interoperable services used beyond their originating context. What such durable infrastructure must additionally do is connect to and draw value from the short-lived, transient activity through which science actually happens. This includes the many experiments, questions, iterations, null results, and the informal exchanges within and across labs that are today filed away or discarded.

To this day, scientific infrastructure is usually seen as a static scaffolding that is built for specific purposes. This results in infrastructure that is often fragmented, overly specific, and short-lived. Typically built by individual labs, such systems suffer from variable quality and limited scalability. Even larger scale governmentally supported infrastructures are often non-interoperable, compartmentalized, and dependent on funding cycles. When they are successful locally, coordination across initiatives, governance models, accessibility, and long-term growth paths remains uneven. This stifles scientific and technological progress, and especially at the cutting edge, where innovation is rapid, infrastructure lags behind.

What is needed is an infrastructure that is intelligent (that can learn, collaborate, decide, and generate) and designed for the dynamic nature of the scientific enterprise. By intelligent I do not mean that every component is an AI model; most of what makes the system intelligent is good engineering of a composable and compatible ecosystem, and the ability to detect its own failures and adapt as science changes. AI is increasingly well positioned to serve this role.

In this perspective, I examine the evolving relationship between science and AI through my own lens of neuroscience. Neuroscience has significant potential for societal impact and is an ideal testbed for a new discovery science[4] paradigm. I assess the gaps in current neuroscience infrastructure that hinder progress and recommend a set of operational guidelines for a new infrastructural paradigm: a scientific ecosystem that changes how we work together to solve our greatest challenges. The single technical capability whose absence now most limits discovery is continuous, machine-actionable compatibility among independently built and governed systems. Supplying it is what converts static infrastructure into a self-correcting scientific instrument and gives AI a transformative rather than incidental role, shifting AI from an analytic tool applied after the fact to an infrastructural capability that maintains the ecosystem itself.

## Neuroscience and AI: A Synergistic Playground

Neuroscience research presents enormous challenges. Each human brain contains about 86 billion neurons and roughly the same number of non-neuronal cells[5], and over 100 trillion connections, so understanding neuroanatomy, molecular dynamics, and neurophysiology across spatial and temporal scales in multiple environments is a monumental task. In turn these inform the study of brain development, behavior, as well as interventions for learning disabilities, neurological conditions, and neuropsychiatric disorders. All this requires observing individuals across a broad range of biological, environmental, developmental, and lifespan dimensions (see Figure 1). Because many of these studies are extremely difficult or currently impossible to perform in humans, much of our current understanding is derived from studies in non-human species.

In terms of infrastructure, these inquiries have led to significant advancements in tools and technologies and have spurred the launch of numerous brain initiatives around the world. These initiatives have aimed at delivering a blueprint for brain function by fostering collaboration, standardizing protocols, and supporting scalable data production and integration. Several large-scale consortial projects have been funded by the US NIH BRAIN Initiative[6] to map cells (BICAN) and circuits (CONNECTS) to behavior (BBQS) and are emblematic of this new era. These consortia are generating comprehensive datasets across species. They required a level of coordination and data integration that needed the creation of an entire neuroinformatics project portfolio and the development of data repositories[7], computational tools, new standards and knowledgebases, and platforms.

These are not the only projects. Efforts like BigBrain[8] and the Fly brain[9] project are mapping the cellular anatomy of various organisms with increasing resolution, and EBRAINS[10] is an EU effort to accelerate collaborative brain research. Initiatives such as ABCD[11], ABIDE[12], OpenNeuro[13], HCP (Human Connectome Project[14]), ENIGMA[15], and ADNI (Alzheimer's Disease Neuroimaging Initiative[16]) have accelerated large-scale human neuroimaging and genetics data collection and sharing within and across nations and labs. Large-scale registries and biobanks, such as the Swedish National Patient Register[17,18] and UK Biobank[19], are increasingly complementing these studies, enabling epidemiological research to be conducted at an unprecedented scale by integrating genomic, imaging, and extensive clinical data from vast populations. Many of these initiatives now rely on computational platforms (e.g., brainlife.io[20], Terra[21]) and controlled research workspaces (e.g., UK Biobank RAP[22]), making repositories only one class of the relevant infrastructure rather than the whole system. As a specific example, brainlife.io provides a concrete precedent for the platform class: its applications expose machine-readable input and output expectations, automatically accept or reject compatible data objects before execution, and preserve provenance across reproducible cloud workflows.

These advances in existing neuroscience infrastructure have resulted in a fragmented landscape of four classes (see Fig 2a). Repositories preserve, curate, and distribute data and metadata, but often expose different interfaces, metadata models, and access paths; computation infrastructure provides scalable execution environments, but access is not consistently federated; knowledgebases and standards organize concepts, schemas, ontologies, and quality expectations, but the multiplicity generates translation and search burdens; and platforms already deliver modular analyses, provenance-aware and reproducible execution, and even automatic checks of whether a data object is compatible with a tool before it runs, but each remains largely a silo, weakly connected to the other classes and to the wider ecosystem. I believe infrastructure-level contracts can make data, compute, workflows, and provenance more directly composable. The next challenge is to make these classes interoperable, auditable, and adaptive participants in an AI-ready ecosystem through explicit contracts for machine-actionable information exchange.

These are neither interchangeable nor failed efforts. Community-owned standards such as BIDS[23] and NWB[24] show that shared formats can achieve broad, cross-tool adoption when paired with validators and repository incentives, yet also that adoption stalls wherever sustained curation and support go unfunded. ENIGMA[15] demonstrated that protocol-harmonized, distributed analysis can pool evidence across dozens of sites and jurisdictions without centralizing raw data, but at the cost of substantial manual coordination for every study. OpenNeuro[13] showed that archives deliver their greatest value when connected to computation. Cloud analysis platforms, in turn, showed that packaging data, reproducible workflows, and compute together can broaden who is able to do computational neuroscience at all. And the trajectory of the INCF[25] illustrates how fragile cross-national coordination becomes when funding and policy priorities shift. The recurring lesson is that isolated excellence does not compose. The same fragmentation is visible prospectively: schemas and terminologies, atlases, workflow systems, and models are being built in parallel across the ecosystem, with significant redundancy. Each such effort would gain from a connected ecosystem in which expertise, prior work, and ongoing activity are discoverable and coordinated as AI advances.

AI holds immense potential to take on this staggering volume and complexity and revolutionize neuroscience by connecting information, transforming computation, and generating new knowledge[26,27]. Over the last decade, the evolution of AI technologies has been rapid, driven by advancements in computational power, data availability, and algorithmic innovation, many inspired by neuroscience. Today's AI landscape is characterized by foundational models, large-scale pre-trained models that can be adapted for a wide range of downstream tasks, offering many generalization capabilities. Reinforcement learning has enabled AI systems to improve representations through trial and error, a process that mirrors aspects of scientific discovery and human learning. Furthermore, "agentic paradigms" are gaining traction, where AI systems are designed to act autonomously, make decisions, and interact with their environments, moving beyond passive analysis to active experimentation. Reasoning capabilities are increasingly being integrated into AI, allowing systems to infer, deduce, and generate hypotheses. Embodied intelligence, where AI systems can perceive and interact with the physical world, also holds immense promise for robotics and experimental control in neuroscience[28]. These capabilities differ sharply in maturity across the ecosystem with many state-of-the-art systems being launched by commercial entities rather than through grassroots efforts.

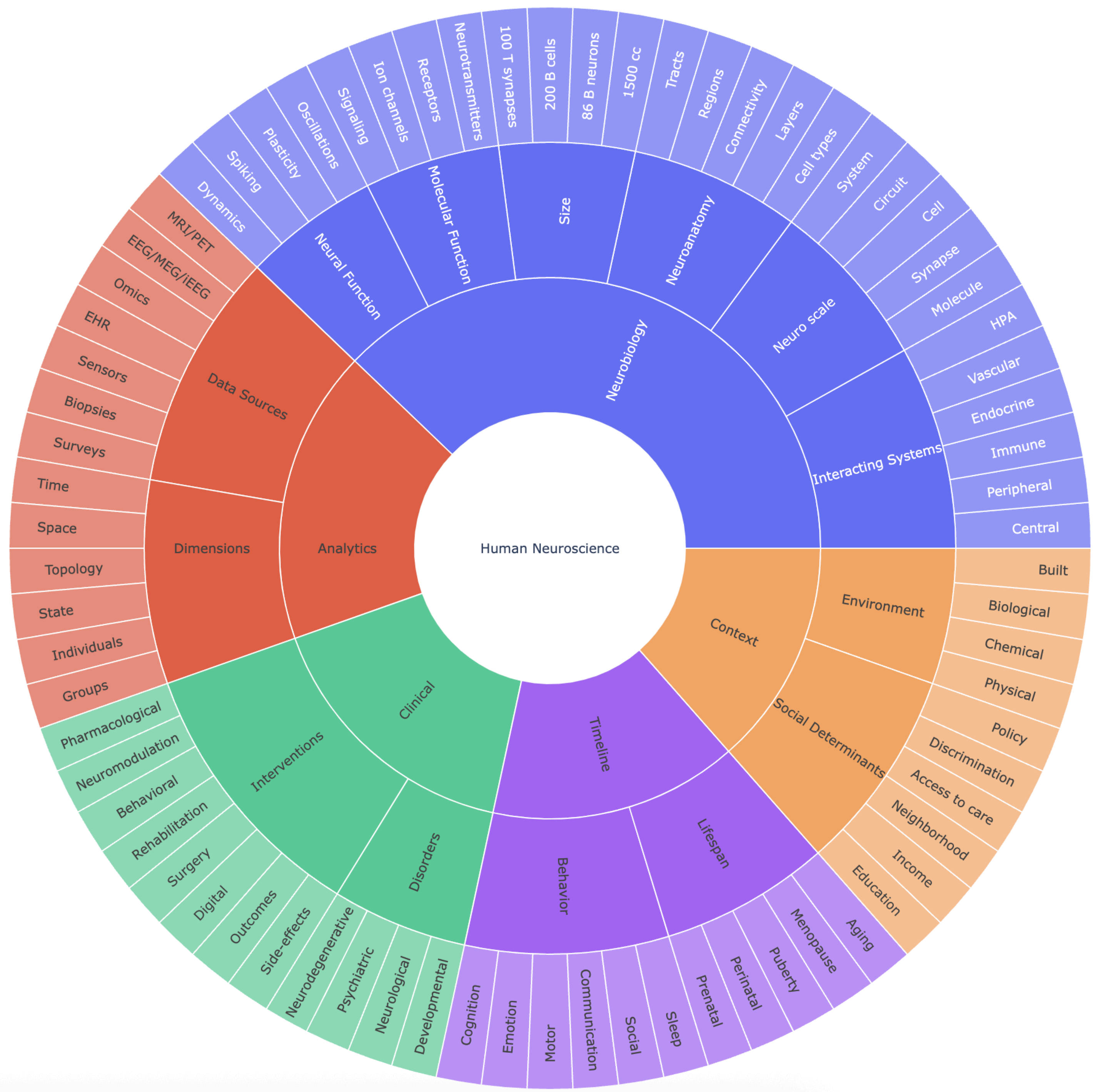

***Figure 1. Complexity of Human Neuroscience.*** *To fully connect knowledge in human neuroscience research, our system will have to assemble data and information across all these elements and connect to the fundamental neuroscience work being conducted in other species. Such a system needs to contextualize any piece of information by its provenance and meaning, which will require rich data and metadata to be available. Further, curating, processing, and connecting such information will require a usable and programmable interface that allows both humans and machines to read and write information. To ensure privacy, such transactions will need to be authorized appropriately depending on the sensitivity of the information. Only an intelligent system in partnership with experts who understand the nuances and failure modes of research can connect this large information space and make it accessible appropriately.*

For all that, AI in and for neuroscience is still in its infancy. Most foundational models for neuroscience[29–33] and similar efforts to understand the brain through AI models[34–36] are highly constrained. For AI to effectively assist in reshaping this field, three conditions must be met. First, comprehensive and reliable information from labs, protocols, data, and publications need to become fully available, similar to the plethora of text that training of large language models (LLMs) relies on. The information needs to span both successes and failures and allow for a comprehensive understanding of the entire data lineage, from pre-acquisition through all processing stages. Second, this rich information needs robust and specialized infrastructure that can support the massive data volumes, facilitate seamless integration of diverse data types, and provide computational resources for the training and deployment of sophisticated AI models. Moreover, it needs to enable rapid cycles of experimentation, data analysis, and model refinement. Third, these interactions within the neuroscientific ecosystem need to produce information and knowledge that is continuously assessed for validity, disseminated broadly, and treated as communication signals in an AI-based system.

Many foundational requirements (persistent identifiers, versioned schemas, provenance, reproducible execution, auditable access, and minimal participation contracts) are not new AI inventions. They are prerequisites that make AI useful and safe in scientific settings. The near-term role of AI in such an ecosystem is to make infrastructure more navigable, computable, and responsive: enriching metadata, mapping related schemas, flagging missing provenance, recommending compatible workflows, summarizing quality signals, and routing users to relevant expertise.

The intersection of neuroscience and AI is therefore a practical stress test for scientific infrastructure. The goal is to move towards a future where AI not only assists but actively participates in the scientific process, designing and even running experiments, interpreting results, and proposing subsequent steps, ultimately driving a more agile and efficient path to neuroscientific understanding[37]. Emerging AI systems already coordinate analyses across repositories, compute environments, knowledgebases, and platforms. More autonomous forms of scientific discovery are already emerging[38,39]. The unresolved barrier is that they are not yet consistently connected through interfaces that allow humans and AI systems to observe state, exchange provenance, detect drift, and adapt the ecosystem as science changes.

Infrastructural ecosystems are critical for accelerating AI for neuroscience and creating synergy across institutions for knowledge discovery, validation, translation, and clinical applications. In the next section I discuss eight challenges that are limiting the transition to a more effective ecosystem for science. These challenges differ in age and tractability, but what is important is that their interaction is where we have the most gaps. Petabyte-scale and emerging exascale workflows, AI-readiness, and the under-recognition of infrastructure labor have turned a set of longstanding, separately managed problems into one systems problem that piecemeal, project-by-project uncoordinated responses will make the ecosystem more brittle.

## Challenges in the Current Neuroscience Ecosystem

### *1. Complexity of Neuroscientific Inquiry*

Neuroscientific inquiry has evolved into a remarkably rich and complex endeavor, encompassing a vast array of sophisticated methodologies and investigative approaches, spanning diverse domains. For example, one can capture in vivo brain activity using technologies like Neuropixels probes that enable the simultaneous recording from hundreds of neurons in awake behaving animals[40] (including humans[41]). As another example, ex vivo examination of human brain tissue can involve advanced tissue-clearing and expansion microscopy techniques[42,43]. These render samples transparent and/or physically enlarged to enable unprecedented observation of gene expression and proteins at cellular and sub-cellular resolution[43]. Such detailed insights are essential for detecting subtle, yet significant, differences between individuals, especially when studying neurodegenerative conditions such as Alzheimer’s Disease and Related Dementias (ADRD).

Each neuroscientific experiment, regardless of its specific focus, demands a highly detailed and multistep protocol, integrating a multitude of advanced technologies. Executing such workflows requires meticulous planning, rigorous implementation, and seamless integration of diverse technological platforms. In many situations these new questions require new instruments, new algorithms to analyze the new data, and new models of the system under study. This makes neuroscience today inherently interdisciplinary: progress depends not only on deep biological expertise but also on mastery of engineering, data science, software development, cloud operations, and increasingly, AI.

### *2. The Data Explosion*

As already touched on, a key challenge is the volume of data. As the neuroscience community pushes the boundaries of what can be observed and quantified, the volume of data generated has surged exponentially. Neuroscience has moved from collecting megabytes (MB) and gigabytes (GB) of data to routinely handling terabytes (TB) and increasingly, petabytes (PB). This scale is driven by the complexity of experiments, the resolution of modern instruments, and the ambition to capture multi-species, multi-scale phenomena across time. Such a dramatic increase in data volume presents both unprecedented opportunities and serious challenges[44]. It necessitates worldwide infrastructures that are not just storage solutions, but end-to-end platforms enabling data curation, provenance tracking, cross-site collaboration, computation, and long-term access.

Many repositories are distributing PBs of data around the world resulting in significant new knowledge from existing datasets, reducing some of the needs for new data collection and its associated costs and resource consumption. However, even knowing what data exists is a challenge, and most primary data are not making their way out of labs in a timely manner, with many PIs and institutions holding on to data that have potential beyond the reasons for which they were generated. Thus, there is an incomplete picture of the data that exists in the world at any point in time, and no way to evaluate whether such data can answer pressing questions.

No widely adopted federated infrastructure yet coordinates data needs and collection, integration, computation, and feedback across these infrastructure classes. A system capable of indexing, accessing, and assessing data at any instant could address many of these challenges. It would also immediately serve the needs for training and validating AI models.

### *3. Fragmentation and Lack of Coordination*

The lack of coordination of common standards or a “lingua franca” for data formats, metadata, analysis pipelines, and sharing protocols severely impedes collaboration and data reuse and increases the potential for errors and miscommunication. As a result, there is great variation in the quality and precision of information across many resources being generated and used. Such imprecision, in turn, influences the reliable training of AI models and the correctness of AI systems.

Despite several advances to consolidate practices across some communities in neuroscience, overall it remains largely a “cottage industry” in which individual labs often develop bespoke protocols and software solutions for specific, localized problems. While this independence fosters creativity and expediency, it undermines interoperability. Data generated in one lab may be incompatible with tools used in another, rendering integration or consolidation difficult or impossible. Such one-off solutions also increase the risk of errors and reduce sustainability when compared to community-maintained, well-engineered software systems.

Standards such as BIDS (Brain Imaging Data Structure[23]), NWB (Neurodata Without Borders[24]), and OME-Zarr[45] have made significant strides in community-driven harmonization of data formats and metadata across domains like neuroimaging, neurophysiology, and microscopy. However, adoption remains inconsistent despite being driven by public data repositories. Incentives for using such standards are limited, and many researchers face steep learning curves or lack institutional support to transition. Further, a lack of institutional or other resources and priorities reduces engagement with the development of these standards, leading to significant lags in the timely deployment and utility of these resources. The efforts of the International Neuroinformatics Coordinating Facility (INCF[25]), an international body originally established as an outcome of the Organisation for Economic Co-operation and Development (OECD), delivered early and coordinated progress in neuroinformatics, but it has since been severely limited in its efforts as a result of policy changes and funding priorities across nations. As a result, many communities remain fragmented today.

### *4. Skill and Knowledge Gaps Persist Across the Ecosystem*

Infrastructure is only as useful as the ability of researchers to effectively utilize it. The use of complex data platforms, computational workflows, and AI models requires specialized expertise in software engineering, data management, and statistical and algorithmic reasoning. However, access to training and technical support is highly uneven. Many labs lack personnel with the necessary computational background, and even where training is available, it may not be tailored to the specific needs of neuroscience. This expertise gap limits the reuse of data, the development of robust pipelines, and the reproducibility of results. Moreover, it creates a dependence on a small number of highly skilled individuals, leading to bottlenecks in collaborative projects and contributing to burnout and attrition.

While the complexity of neuroscience has increased, federal support for training programs has reduced, especially in the US. This has reduced broad baseline knowledge and skills that can contribute to career stability and resilience. As AI systems are increasingly used, fundamental questions arise around the role of education and the future of work. Training itself needs to reflect these changing directions, but is not adapting quickly to the current nature of technological changes. Many individuals are turning to AI systems for education, without a clear consideration of the ability or inability of these systems to deliver accurate knowledge. Thus, there is a significant potential for an even more brittle workforce and scientific ecosystem.

### *5. Infrastructure Inequities*

One of the most pressing issues is the striking variability in infrastructural resources across labs, institutions, and countries. In contrast to those in under-resourced settings, researchers in well-funded institutions in high-income countries often enjoy access to state-of-the-art equipment, dedicated IT support, scalable storage, and high-bandwidth networks. This imbalance limits participation in large-scale collaborative science and introduces systemic challenges in who can contribute to and benefit from cutting-edge research. It also poses significant obstacles to building globally representative datasets. Under-resourced groups often have to build local adaptations to meet immediate needs. Without mechanisms to support, generalize, and connect those efforts, valuable local expertise remains invisible, underfunded, and difficult to integrate into shared infrastructure. Access to data alone does not create equitable participation. Repositories lower barriers when they are connected to usable platforms, trusted or controlled computation environments, documentation, training, visualization, and reproducible workflows; otherwise they may reproduce skill, compute, and support inequities. Cloud analysis platforms[20–22] have already shown the difference: by pairing standardized datasets with hosted, reproducible workflows and visualization, they have enabled researchers who lack local pipelines, compute, or engineering support to run analyses and produce results that would otherwise be out of reach. However, geopolitical and financial circumstances severely limit which datasets are available and who gets access. A dynamic infrastructure should therefore measure not only whether resources are available, but also whether researchers can access them, operate in more collaborative forms, and understand provenance both within and across the ecosystem.

### *6. Ethical and Regulatory Fragmentation*

In addition to technical and resource disparities, neuroscientific infrastructure must navigate complex ethical and regulatory landscapes. Data privacy, security, and consent, particularly for human data, are governed by diverse frameworks and policies that vary by country and institution. These discrepancies can complicate international collaborations, hinder data sharing, and delay scientific progress. Given some of the health-related challenges neuroscientists are trying to solve, the collection and exchange of information across the world's population is an essential requirement for the development of robust treatments. We have already seen inconsistencies in diagnosis of mental health disorders and selection of treatments on a trial and error basis. As AI systems begin to interact more deeply with neuroscientific datasets in the march towards precision medicine, these ethical concerns will only grow. AI models trained on poorly governed data may perpetuate biases or expose sensitive information.

### *7. Biased Attribution*

Neuroscience is a team science. Every advance stems from the dedication and contributions of many individuals. However, in the contemporary landscape of academia and industry, the relentless pursuit of limited tenure-track positions, grant funding, and competitive industry roles has fostered an environment where quantity often overshadows quality. Individual credit is derived through publications. The "publish or perish" mentality incentivizes researchers to produce a high volume of papers, sometimes leading to fragmented research, minimal new insights, or rushed submissions. The pressure to publish often means that the development of robust, well-documented, and easily usable software takes a backseat. The individuals and teams responsible for building and sustaining infrastructural and software elements often receive inadequate recognition[46], leading to lack of maintenance of these elements and hindering reproducibility. Such trends highlight a systemic issue where the current reward structures often fail to adequately recognize the diverse forms of contribution that are essential for scientific advancement.

### *8. Infrastructure is Underfunded*

The foundational infrastructure crucial for scientific progress, particularly in fields like neuroscience, is severely underfunded, creating a severe bottleneck for innovation and scalability. A portion of infrastructure is usually supported through indirect costs associated with institutional components. However, a vast and increasingly critical segment of neuroscience infrastructure - encompassing data coordinating and dissemination centers, sophisticated data repositories, and the essential work of software development and engineering[47], does not receive funding commensurate with the escalating needs of scientific projects. This disproportionate funding model prevents these vital components from scaling effectively[48] to meet the demands of modern research.

While proposals to address this disparity have emerged, notably within the European Union, they have yet to translate into concrete policy changes and sustained funding mechanisms, which are themselves under pressure[49]. The development and deployment of sophisticated AI models demand substantial investments in computational power, extensive data storage, and specialized engineering expertise. Both funds and expertise are increasingly moving to large technology companies, where commercial incentives are often not aligned with long-horizon fundamental research. Without a significant recalibration of funding priorities, the burgeoning potential of AI in scientific discovery will be severely hampered by inadequate infrastructure, widening the gap between scientific ambition and practical execution.

| Principle | Primary focus | Who it targets | Relation to perspective |
|---|---|---|---|
| **FAIR**<br>**F**indable<br>**A**ccessible<br>**I**nteroperable<br>**R**eusable | "ability of machines to automatically find and use the data, in addition to supporting its reuse by individuals." | Data producers & users (human + machine) | All components of infrastructure should operate under FAIR considerations. |
| **CARE**<br>**C**ollective Benefit<br>**A**uthority to Control<br>**R**esponsibility<br>**E**thics | "encouraging open and other data movements to consider both people and purpose in their advocacy and pursuits." | Indigenous Peoples and communities and their data users | All components of an infrastructure should adhere to these CARE principles including AI systems, paying attention to data sources. |
| **TRUST**<br>**T**ransparency<br>**R**esponsibility<br>**U**ser focus<br>**S**ustainability<br>**T**echnology | "remind data repository stakeholders of the need to develop and maintain the infrastructure to foster continuing stewardship of data and enable future use of their data holdings" | Repository operators & their users | All components of an infrastructure need to provide TRUST through communal governance and stewardship. |

**Box 1. The FAIR, CARE, and TRUST principles.** These principles provide a foundation of values that should apply to any infrastructure, especially an AI-embedded one. As much as these principles define guidelines for data management, Indigenous Peoples' data governance, and digital repositories, the recommendations in this perspective help operationalize these guidelines towards implementation in an intelligent infrastructure for science.

Sources: FAIR: https://www.nature.com/articles/sdata201618, CARE: https://www.gida-global.org/care, TRUST: https://www.nature.com/articles/s41597-020-0486-7

## Recommendations for an AI-Aligned Neuroscience Infrastructure

Because these challenges compound rather than stack, the neuroscientific community should build coordination layers over existing efforts and re-engineer the interfaces among them to support a more seamless and collaborative scientific process. Over the past two decades, my work in neuroscience has involved leading numerous initiatives to develop standards, software, and infrastructure. This extensive experience has provided me with ample opportunities to observe, reflect on, and evaluate the impact of these efforts, and to interact with many researchers throughout the process.

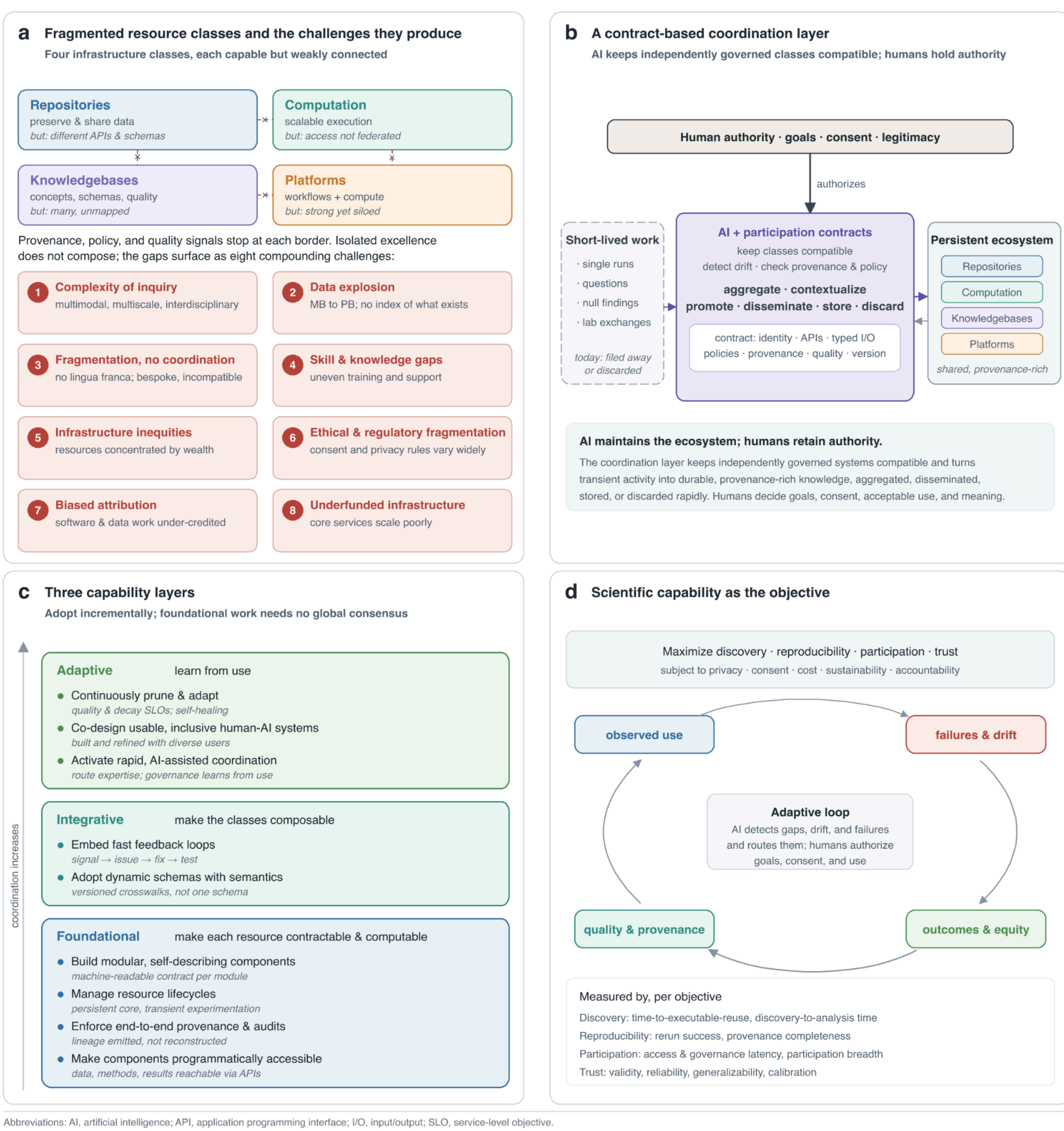

*Figure 2. Intelligent infrastructure as a dynamic coordination layer. a, Today, repositories, computation, knowledgebases, and platforms each do capable work but are not machine-actionably connected; provenance, policy, and quality signals stop at each border and the exploration behind results is discarded, so discovery and reuse take months of manual coordination. b, The proposed shift adds a coordination layer in which each resource publishes a machine-readable participation contract; AI helps detect incompatibilities, propose repairs, and route decisions to accountable stewards, and promotes short-lived, transient activity into persistent, provenance-rich knowledge, while human authority governs goals, consent, and legitimacy. c, The capabilities are organized as three layers (foundational, integrative, and adaptive) comprising design patterns that can be adopted incrementally without global consensus. d, The ecosystem optimizes scientific capability (discovery, reproducibility, participation, and trust, under privacy, consent, and sustainability constraints) through an adaptive loop in which AI detects failures and drift and humans authorize goals and use.*

In many of these interactions, the FAIR[50], CARE[51], and TRUST[52] principles (see Box 1) frequently emerge as invaluable. The FAIR (Findable, Accessible, Interoperable, and Reusable) principles for data are widely recognized as the global standard for effective data management. These are complemented by the CARE (Collective benefit, Authority to control, Responsibility to providers, and Ethics) principles for Indigenous Peoples' data governance. Although CARE was developed in that context and should not be generalized in a way that erases it, its emphasis on collective benefit, authority, responsibility, and ethics highlights governance obligations that AI-ready infrastructures must represent explicitly. More recently, the TRUST (Transparency, Responsibility to users, User focus, Sustainability, and Technology) principles were developed for improving digital repositories. But, whereas these principles establish an important foundation for scientific values that broadly apply to any digital infrastructure, their operationalization remains severely haphazard. The varied approaches to implementing these principles have, in fact, led to many of the challenges highlighted previously. Operationalizing FAIR, CARE, and TRUST requires attention to tensions among them across the full infrastructure ecosystem. Data that are maximally findable and reusable may not be ethically reusable in every context. Consent, withdrawal, community authority, jurisdictional rules, and privacy protections should be represented as machine-actionable temporal constraints across repositories, computation infrastructure, knowledgebases, and platforms. The legitimacy of those constraints remains a human and community responsibility, while their implementation should adapt as the system exposes failures, inequities, and new scientific needs.

Based on my observations and interactions, and building on values encoded in the FAIR, CARE, and TRUST principles, I outline operational recommendations to guide the transformation towards a more dynamic, self-learning, and self-correcting ecosystem in which to do our scientific work. Foundational capabilities make individual resources contractable and computable. Integrative capabilities connect repositories, computation infrastructure, knowledgebases, and platforms through registries, semantic crosswalks, workflow interfaces, provenance exchange, quality signals, and policy-bound computation. Adaptive capabilities then operate over this connected substrate: governance, AI-assisted coordination, schema evolution, deprecation, expert routing, incentives, and quality review should be informed by observed use, failures, scientific outcomes, and community priorities. This avoids making broad governance consensus a prerequisite for progress while ensuring that the system remains accountable, corrigible, and responsive as it grows. These requirements are composed into design patterns for a federated, AI-ready ecosystem (Figure 2). Box 2 clarifies these capabilities as nine design patterns, each with a concrete first step and the scientific capability and metric by which to judge it. The objective function is scientific capability: faster discovery and reuse, better reproducibility, broader participation, lower coordination cost, and trustworthy AI-assisted science under constraints of privacy, consent, cost, sustainability, and accountability.

### A. Foundational capabilities

#### *i) Build Modular, Self-Describing Components*

Every module (e.g., a data pipeline, AI model, or service) should publish a machine-readable contract: its typed inputs, operations, outputs, accepted schemas, versions, and access constraints. That contract lets another system, or an AI agent, determine compatibility and validate behavior without human mediation, making modularity an enforceable obligation of participation rather than a design preference. This design promotes cohesion and agility, accelerated experimentation, scalability, and obviates the need to re-architect systems from scratch. To avoid confusion, inefficiency, and resource wastage from uncoordinated, redundant, or overlapping modules, systems should analyze the downstream impact of updates and identify underused or outdated modules for deprecation. Governance must enable broad participation in module addition while upholding strict quality and integration standards for long-term resilience. Modules should also avoid tight coupling, which can lead to unpredictable cascading changes or an inability to accommodate evolving standards, resulting in degraded performance or loss of interoperability. The aim is to ensure robustness without compromising stability, allowing for rapid iteration while maintaining a stable yet evolving system. Each module then functions as a computational unit that plugs into both human and machine feedback loops.

#### *ii) Manage Resource Lifecycles (Persistent vs. Transient)*

Distinguishing persistent from transient resources is what lets one system support both rapid experimentation and durable reproducibility. This matters for any AI coordination and development, since models rely on well-defined data lineage and stable core interfaces yet must stay responsive to the cutting edge of knowledge. Persistent identifiers and stable core interfaces preserve what must remain reproducible, while transient runs and intermediate outputs carry their intended lifespans and expire once their usefulness ends rather than accumulate unmonitored.

**Box 2. Design patterns, implementation routes, AI capabilities, and metrics**

| Design pattern | Implementation | AI role and scientific capability metric |
|---|---|---|
| **Foundational: make each resource contractable and computable** | | |
| **Build modular, self-describing components** | Publish a machine-readable contract for each module (data pipeline, model, instrument, or service): persistent identity, steward, object types, accepted schemas, typed inputs and outputs, versions, APIs, access constraints, and provenance and quality requirements. | AI validates contracts, detects missing fields, flags incompatible inputs, and suggests compatible resources.<br>Metrics: contract coverage, conformance rate, time to first successful reuse. |
| **Manage resource lifecycles (persistent vs. transient)** | Tag every artifact with an intended lifespan; keep persistent identifiers and core interfaces stable while allowing rapid transient experimentation; audit archival completeness and expire or clean up temporary data. | AI tracks lineage and lifespan, flags orphaned or stale artifacts, and proposes promotion to persistence or expiry.<br>Metrics: lifecycle coverage, stale-artifact rate, core-interface stability. |
| **Enforce end-to-end provenance and audits** | Emit a provenance event stream continuously (acquisition, curation, schema conversion, workflow execution, model invocation, access decision, quality review, publication, deprecation) rather than reconstructing lineage at publication. | AI detects missing lineage, anomalous transformations, irreproducible branches, and attribution gaps.<br>Metrics: event completeness, audit time, error-recovery time. |
| **Make all components programmatically accessible** | Expose data, methods, protocols, results, and decisions through APIs; package each core workflow as a portable, inspectable object (code, container, parameters, tests, quality checks, provenance emission) able to travel across platforms and trusted workspaces. | AI recommends workflows, predicts execution failures, explains quality-control outputs, and summarizes provenance.<br>Metrics: programmability, rerun success rate, failed-run diagnosis time. |
| **Integrative: make the classes composable** | | |
| **Embed fast feedback loops** | Emit machine-actionable signals (quality flags, reproducibility failures, schema conflicts, access denials) and route each to an owner so it drives an issue, a change, and a confirming test, closing the loop across systems. | AI clusters signals, routes them, and drafts fixes for human approval.<br>Metrics: signal-to-resolution time, recurring-failure reduction. |
| **Adopt dynamic schemas with semantics** | Maintain versioned semantic crosswalks among repository schemas, standards, ontologies, atlases, and knowledgebases; record confidence, worked examples, unresolved conflicts, and expert decisions so federation proceeds under disagreement. | AI proposes mappings, detects drift, clusters equivalent concepts, and routes ambiguous mappings to experts.<br>Metrics: mapping coverage, unresolved-conflict rate, discovery-to-analysis time. |
| **Adaptive: learn from use** | | |
| **Continuously prune and adapt (self-healing ops)** | Treat scientific quality and semantic decay as service-level objectives with owners and deadlines; regularly review, correct, and prune outdated, low-quality, or redundant components while evolving with new data and techniques. | AI clusters recurring failures, predicts technical debt, recommends repair or deprecation, and triages support.<br>Metrics: quality-flag resolution time, recurring-failure reduction, support latency. |
| **Co-design usable, inclusive human-AI systems** | Co-design interfaces and workflows with users of diverse skills, backgrounds, and abilities; refine continuously from real use-cases, including AI-driven conversational interaction, so the system serves the whole community. | AI-driven interaction resolves many issues directly and surfaces usability barriers.<br>Metrics: task completion, participation breadth, support latency. |
| **Activate rapid, AI-assisted coordination** | Coordinate across the ecosystem: route questions to relevant expertise, run policy-bound computation next to sensitive data, and revise governance from observed access conflicts, failures, reuse, equity indicators, and scientific outcomes. | AI summarizes signals, routes expertise, prechecks access constraints, and proposes contract or policy updates while humans retain legitimacy and accountability.<br>Metrics: coordination latency, governance latency, secure reuse rate, inclusion. |

### *iii) Enforce End-to-End Provenance & Audits*

Scientific trust is severely undermined when the origins and transformations of data and models are undocumented or lost, leading to irreproducible results, misattribution, or undetected misuse. Every data element, model, and decision must be traceable from origin to current state through robust provenance and auditability. This needs a transparent record of creation, evolution, and trustworthiness. This is especially critical for intelligent systems, as audit trails enable reproducibility and validation, fostering trust in AI outputs and allowing systems to self-diagnose or explain their reasoning. For humans, it facilitates crucial evaluation and correction, and perhaps more importantly attribution, especially in a joint world of human and AI system interaction.

### *iv) Make All Digital Components Accessible Programmatically*

For AI to collaborate and for results to reproduce, scientific work must be machine-actionable, not merely human-readable. Data, methods, protocols, results, and decisions must all be reachable programmatically. This fundamental shift is critical because, to integrate AI as a scientific collaborator, the system must be fully programmable, capable of simulating hypotheses, running experiments, adapting to new knowledge, and learning autonomously, thereby enabling self-adjusting, AI-augmented science. Recent AI co-scientists[38,39,53] are examples of what such automation and accessibility enables.

## B. Integrative capabilities

### *i) Embed Fast Feedback Loops*

A feedback-responsive system is fundamental to scientific infrastructure, as it supports short feedback cycles among data collection, analysis, interpretation, and system refinement, ensuring agility and responsiveness to emerging needs and discoveries. Software infrastructures close the loop routinely: user feedback becomes an issue, an issue drives a change, and an automated test confirms the fix. Science needs the same loop, but its signals (quality flags, reproducibility failures, schema conflicts, access denials) must be emitted in machine-actionable form so they propagate across independently governed systems rather than staying trapped in one platform leading to risk of issues persisting unnoticed, leading to stagnation or reduced trust.

### *ii) Adopt Dynamic Schemas With Semantics*

Rather than wait for a universal schema, the ecosystem should maintain versioned crosswalks among the schemas that already exist, recording confidence, worked examples, and unresolved conflicts, so federation can proceed under disagreement instead of being blocked by it. Structured, semantic representations let AI systems interpret, reason, and act across domains, and let humans evaluate what those systems produce or align. The goal of improving shared understanding should be built into the design and operation of future systems.

Such schemas must incorporate new concepts as readily as they expose and refine older ones as understanding changes. If every data element in this ecosystem was structured with a machine-readable description and its provenance[54], it would solve most of the common challenges that are currently encountered in schema development, and would enable dynamic schemas and seamless data movement. Any prior information should be checked for validity against new schemas, identifying elements invalidated by the new or refined scientific models. This proactive approach tackles the risks of opaque and difficult-to-interpret data when clear, shared definitions are lacking or evolve without synchronization. It directly counters schema drift, where ontologies and models evolve inconsistently, preventing broken interoperability and confusion, particularly for AI systems that rely on semantic precision. Furthermore, by making metadata addition intuitive and integrated, we mitigate the risk of sparse, ambiguous, or manually intensive metadata that researchers might bypass, leading to low-quality annotations and fragmented datasets.

## C. Adaptive capabilities

### *i) Continuously Prune & Adapt (Self-Healing Ops)*

Systems must regularly review, correct, and prune outdated, low-quality, or redundant components. This needs to happen across instruments, data, algorithms, and knowledge, while evolving with new data and techniques at the pace of science. This matters especially for AI, which depends on high-quality inputs and must adapt to new contexts without constant human micromanagement. The system must also be able to discard or contextualize changes in knowledge, which means that it should be capable of reasoning about why some knowledge no longer holds or why our understanding has changed.

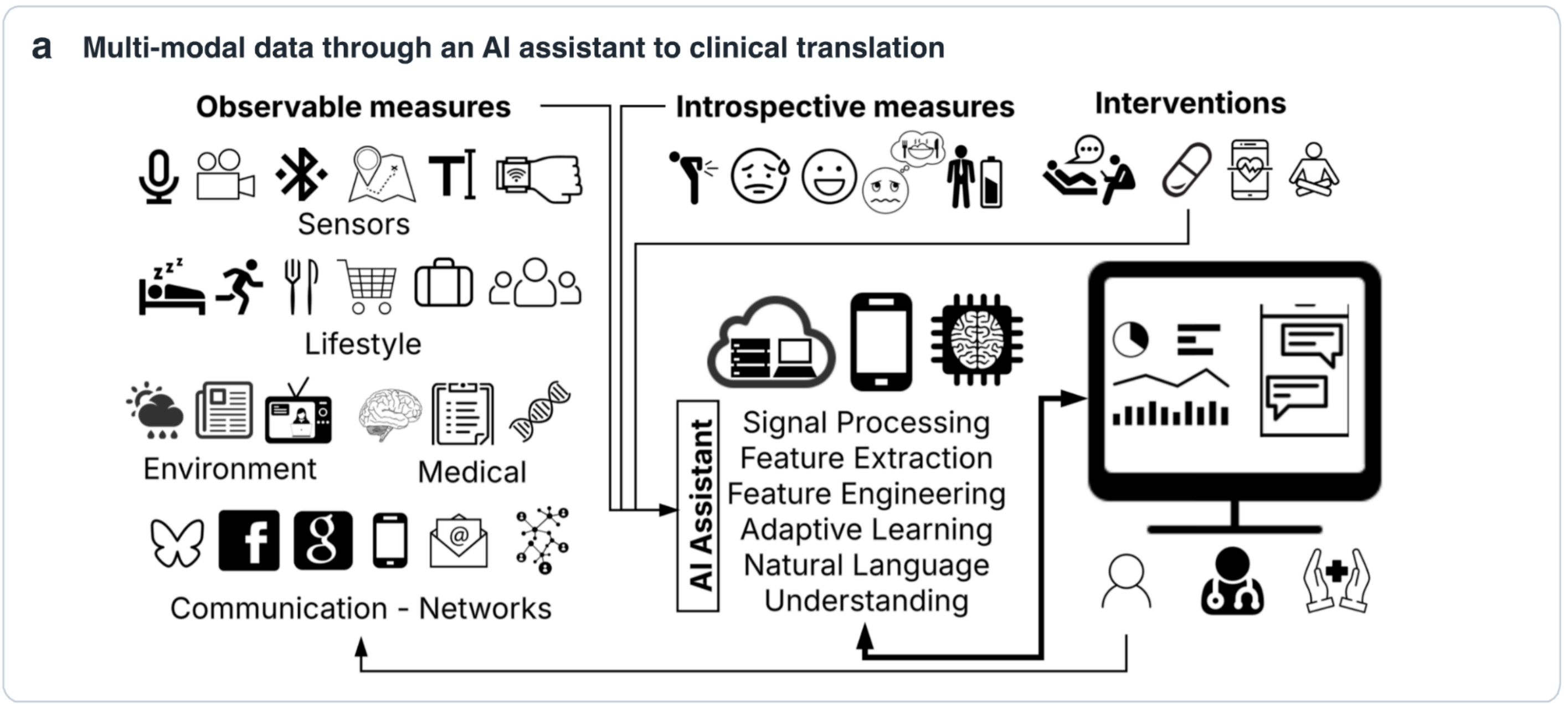

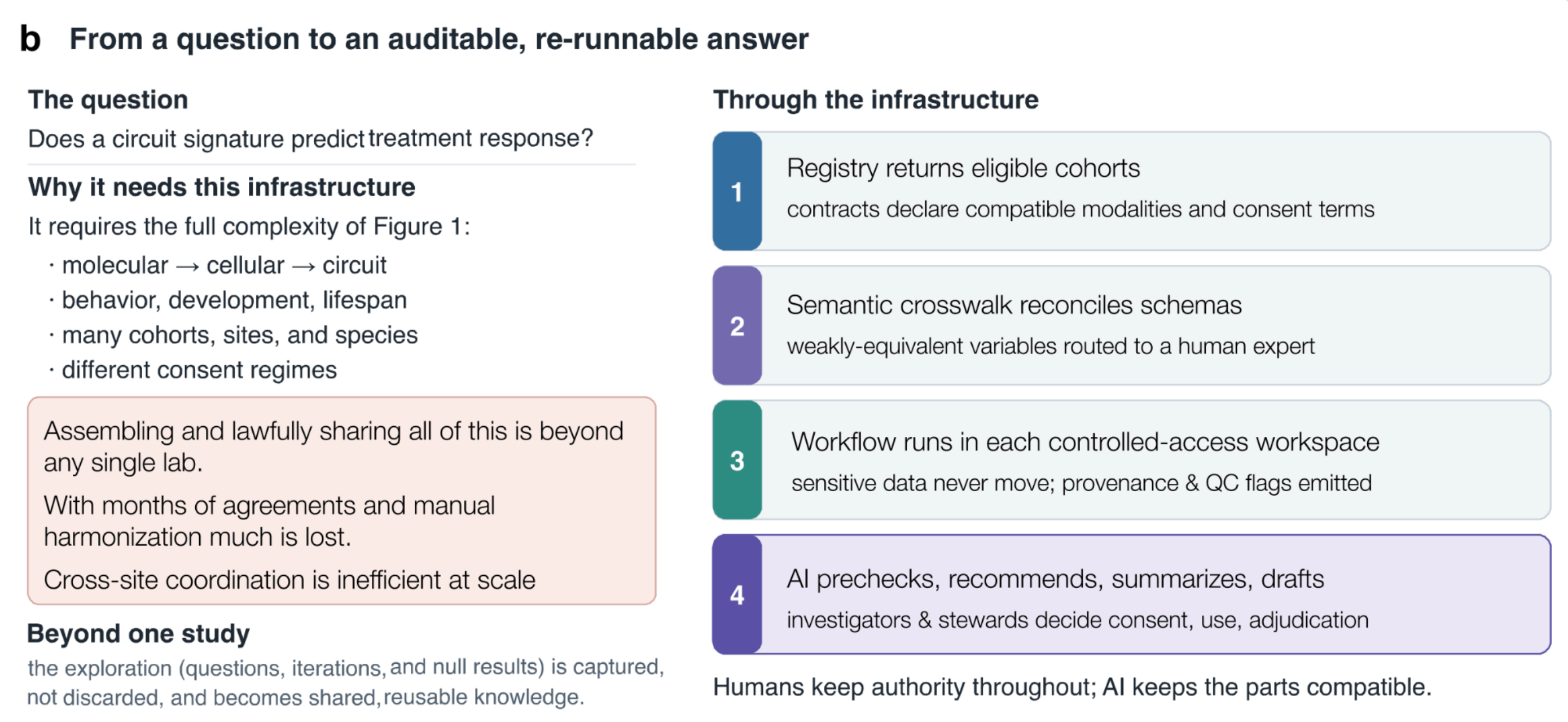

*Figure 3. Precision psychiatry, a use case:. a, The data ecosystem the question requires: multi-modal data from observable measures (sensors, lifestyle, environment, medical records, communication), introspective measures, and interventions flow through an AI assistant (signal processing, feature extraction, adaptive learning, natural-language understanding) to clinical analytics and decision support and to participation by patients, clinicians, and care teams, with continuous feedback. b, Turning questions into an auditable answers: because assembling such breadth across cohorts, sites, and consent regimes is beyond any single lab, a registry returns eligible cohorts, a semantic crosswalk reconciles their schemas (routing ambiguous variables to a human expert), a policy-bound workflow runs beside the data so sensitive data never move, and AI drafts the synthesis while humans retain authority. This turns months of agreements and manual harmonization work into a days-long, auditable, re-runnable analysis in which the exploration behind results is captured rather than discarded.*

### *ii) Co-Design Usable, Inclusive Human-AI Systems*

Systems must serve real people with varying backgrounds, expertise, and goals, not only technical experts but the whole scientific community. Interfaces and workflows should be co-designed with these users and refined continuously from their use-cases, a process that AI-driven interaction now makes far easier to sustain, resolving many issues directly through conversation with autonomous agents. These usage patterns themselves serve as training signals for these systems. Thus, reinforcemental signals should come from both humans and machines across the ecosystem.

***iii) Activate Rapid, AI-Assisted Coordination***

Fragmented knowledge slows everything: when relevant expertise is not at hand, decisions stall, effort is duplicated, and outcomes suffer. Rapid coordination counters this by combining collective intelligence with AI-driven knowledge integration to put critical insights within immediate reach. Coordinating the components of this infrastructure calls for a functional layer that identifies and engages subject-matter experts across fields and supports real-time collaboration and knowledge sharing. AI systems should be integrated to capture and organize distributed knowledge including mistakes, null findings, and practical solutions, making it searchable and easily retrievable. Collaborative tools must exist that facilitate identifying commonalities and divergences in expert opinions (e.g. platforms like Polis[55]). This proactive approach directly counters the risks of inefficiency and missed opportunities that arise from disconnected knowledge. It ensures efficient decision-making and fosters innovation by integrating diverse perspectives.

This coordination is an adaptive layer over human-defined governance, standards, consent constraints, and quality expectations, and governance itself becomes adaptive: access logs, provenance records, workflow failures, schema conflicts, quality signals, user feedback, and scientific outcomes reveal where policies, contracts, standards, incentives, or support structures need revision. AI can detect and route these signals, but legitimacy, acceptable use, consent, and accountability remain human and community responsibilities.

## Implementing recommendations through a viable pilot

To make this concrete, consider a precision-psychiatry question pursued across the ecosystem of sensors and interventions illustrated in Figure 3a: does a specific circuit signature predict treatment response for a given disorder? This can build off the extensive manual coordination that the ENIGMA project currently performs. A Neurobagel[56] or other sample registry could return three eligible cohorts whose contracts declare compatible data modalities and consent terms; a semantic crosswalk reconciles their differing phenotypic data elements and schemas and flags variables that are only weakly equivalent, routing them to a human expert for adjudication. A workflow carrying its container, parameters, and quality checks runs inside each cohort's controlled-access workspace, so sensitive data never move, and every run emits a provenance event stream and quality control flags. AI prechecks access constraints, recommends the compliant workflow, summarizes the quality outputs, retrieves relevant literature and caveats, and drafts the cross-cohort synthesis and analysis, while decisions about consent scope, acceptable use, and the scientific adjudication stay with the investigators and stewards. What today takes months of bespoke data-use agreements and manual harmonization becomes a days-long, auditable, and re-runnable analysis, fitting into emerging co-scientist and agentic workflow systems (e.g. Claude Science[53], ASTA[57], Kosmos[58]).

Such a minimal pilot would turn the proposed patterns into actions. The pilot would require participation contracts, workflows, provenance event streams, a semantic crosswalk, policy-bound execution where sensitive data are involved, quality/decay service levels, and a governance dashboard. Success would be measured by time from discovery to execution, provenance completeness, workflow rerun success, quality-flag resolution, access-policy conflict detection, support latency, and participation by users with different levels of local infrastructure. Funding programs organized around closed-loop research could support a coordinated effort around specific scientific questions from research laboratories, intelligent co-scientist systems, and a marketplace of experimental protocol execution laboratories that feed data and/or directional signals back into the system (e.g., the ARPA-H program IGOR[59]). Sequencing matters because feasibility is uneven: foundational capabilities are an engineering and incentive problem individual groups can begin now, whereas cross-jurisdictional integration and shared governance are constrained by funding, institutional risk tolerance, data-sovereignty law, and a geopolitical climate in which international coordination is increasingly contested. A bounded pilot can show compounding value before that harder alignment is attempted.

## Conclusion: A Call For Change

Neuroscience, with its multimodal and multiscale data, intricate methods that probe multiple species, and profound and urgent societal implications, is both the stress test and the proving ground for a new model: an intelligent, AI-aligned ecosystem that learns from use, coordinates efforts across communities, and continually improves itself.

The recommendations outlined here are necessary conditions for an infrastructure of the future in which humans and AI collaborate productively. The costs of staying with today's fragmented, brittle infrastructure are already here. We waste energy and resources, both computation and human effort, by duplicating tools and infrastructure and regenerating and curating data that should have been FAIR from the start. We open channels for misinformation and disinformation when provenance is weak, incentives reward volume over verification, and many scientific results cannot be reproduced or audited. We widen inequities when only a few institutions can afford to participate at scale, and we drain talent when software and data work remain under-recognized and underfunded. These are not abstract risks; they are daily inhibitors of scientific progress.

I suspect that many of my colleagues agree with the challenges and potential solutions proposed here and the need for tackling them holistically. But almost none are in a position to enact this scale of simultaneous change, and many are not in a position to augment their current research portfolios to take on these challenges. Some would argue we need evidence of impact before scaling, we need to carry out efforts aligned with our limited resources, and perhaps neuroscience does not need such a globally coordinated infrastructure. To them I say that the absence of such coordinated infrastructure limits both the questions we can ask and the answers we can trust, and that a practical path forward is within reach.

Funders can create long-horizon, renewable infrastructure lines, treating data coordination centers, software engineering, and compute commons as core scientific instrumentation, and perhaps most importantly providing support for coordination across global efforts. Institutions and journals can require machine-actionable data and metadata, registered analysis pipelines, and reproducible computation environments, while crediting software, data curation, and infrastructure leadership as first-class scholarly outputs. Consortia can converge on shared interfaces and ontologies, with governance that allows rapid evolution as needs change. Training programs can close skill gaps by pairing domain science with engineering and AI literacy[60], and by valuing negative results and careful curation. And globally, we can lower barriers for under-resourced communities by investing in shared platforms, harmonized ethics frameworks, and equitable access to compute and storage. This increased representation is critical for an AI-enabled ecosystem.

The destination is infrastructure that is reliable enough to fade into routine use yet observable enough to detect anomalies, route resources, coordinate responses, and repair itself when something goes wrong. Done well, this makes AI a dependable participant in maintaining the ecosystem rather than an oracle over it. Building it will require a conscious and deliberative effort towards collective design, sustained investment, and new reward structures. The payoff is a science enterprise that is more rigorous, more inclusive, and dramatically more capable of turning data into understanding and understanding into better lives. If we treat intelligent infrastructure as a central instrument of discovery, not an afterthought, we can meet the complexity of brain science head-on and establish a blueprint other fields can adopt. The above is a plea for a science that finally composes independently governed systems, held in continuous compatibility, letting us ask and answer questions that no single lab, dataset, or model could reach alone. This is not just a matter of more efficiency, but a step toward building a scientific process that is more precise, comprehensive and, ultimately, trustworthy.

## Acknowledgments

This perspective is an outcome of many discussions with many members of the neuroscience community over the last two decades. I thank each of them for their contributions to my evolving thought process. I would also like to thank my wife Katrien Vander Straeten for her patience (in general), numerous discussions, and for editorial work in preparation of this manuscript. I thank Arno Klein, Shoaib Mufti, Kris Ganjam, and Alyssa Picchini Schaffer for several recent discussions related to this perspective. My group's research is supported by NIH R24MH117295, NIH U24DA064429, NIH P41EB019936, NIH U24MH130918, NIH UM1MH134812, NIH UM1NS132358, NIH OT2OD032720, Simons Foundation, and the Lann and Chris Wohrle Psychiatry Fund at the McGovern Institute for Brain Research at MIT. Each of these projects has provided insights about the breadth and depth of infrastructure needed to do scientific work, and the myriad challenges in both constructing and maintaining infrastructure. As such I am indebted to the many colleagues I work with on a regular basis in these projects.

**References**

1. van Laak, D. *Lifelines of Our Society*. (MIT Press, Cambridge, MA, USA, 2023).

2. Güntsch, A. *et al.* National biodiversity data infrastructures: ten essential functions for science, policy, and practice. *Bioscience* **75**, 139–151 (2025).

3. UNESCO. *UNESCO Recommendation on Open Science*. https://doi.org/10.54677/MNMH8546 (2021) doi:10.54677/mnmh8546.

4. Wikipedia contributors. Discovery science. *Wikipedia, The Free Encyclopedia* https://en.wikipedia.org/wiki/Discovery_science (2025).

5. Azevedo, F. A. C. *et al.* Equal numbers of neuronal and nonneuronal cells make the human brain an isometrically scaled-up primate brain. *J. Comp. Neurol.* **513**, 532–541 (2009).

6. Ngai, J. BRAIN @ 10: A decade of innovation. *Neuron* **112**, 3003–3006 (2024).

7. Iyer, S. *et al.* The BRAIN Initiative data-sharing ecosystem: Characteristics, challenges, benefits, and opportunities. *Elife* **13**, (2024).

8. Amunts, K. *et al.* BigBrain: an ultrahigh-resolution 3D human brain model. *Science* **340**, 1472–1475 (2013).

9. Dorkenwald, S. *et al.* Neuronal wiring diagram of an adult brain. *Nature* **634**, 124–138 (2024).

10. EBRAINS. EBRAINS RI - Europe's Digital Infrastructure for Brain Research. *EBRAINS* https://ebrains.eu/.

11. Casey, B. J. *et al.* The Adolescent Brain Cognitive Development (ABCD) study: Imaging acquisition across 21 sites. *Dev. Cogn. Neurosci.* **32**, 43–54 (2018).

12. Di Martino, A. *et al.* The autism brain imaging data exchange: towards a large-scale evaluation of the intrinsic brain architecture in autism. *Mol. Psychiatry* **19**, 659–667 (2014).

13. Markiewicz, C. J. *et al.* The OpenNeuro resource for sharing of neuroscience data. *Elife* **10**, (2021).

14. Van Essen, D. C. *et al.* The WU-Minn Human Connectome Project: an overview. *Neuroimage* **80**, 62–79 (2013).

15. Thompson, P. M. *et al.* The ENIGMA Consortium: large-scale collaborative analyses of neuroimaging and genetic data. *Brain Imaging Behav.* **8**, 153–182 (2014).

16. Weiner, M. W. *et al.* The Alzheimer's Disease Neuroimaging Initiative: a review of papers published since its inception. *Alzheimers. Dement.* **9**, e111-94 (2013).

17. Olén, O. & H Everhov, Å. An updated review of The Swedish National Patient Register as a data source.

*Lakartidningen* **122**, (2025).

18. Everhov, Å. H. *et al.* Diagnostic accuracy in the Swedish national patient register: a review including diagnoses in the outpatient register. *Eur. J. Epidemiol.* **40**, 359–369 (2025).
19. Bycroft, C. *et al.* The UK Biobank resource with deep phenotyping and genomic data. *Nature* **562**, 203–209 (2018).
20. Hayashi, S. *et al.* Brainlife.Io: A decentralized and open-source cloud platform to support neuroscience research. *Nat. Methods* **21**, 809–813 (2024).
21. Science at Scale. *Terra* https://terra.bio/ (2023).
22. Research Analysis Platform. *UK Biobank* https://www.ukbiobank.ac.uk/use-our-data/research-analysis-platform/ (2024).
23. Gorgolewski, K. J. *et al.* The brain imaging data structure, a format for organizing and describing outputs of neuroimaging experiments. *Sci Data* **3**, 160044 (2016).
24. Rübel, O. *et al.* The Neurodata Without Borders ecosystem for neurophysiological data science. *Elife* **11**, e78362 (2022).
25. Abrams, M. B. *et al.* A standards organization for open and FAIR neuroscience: The International Neuroinformatics Coordinating Facility. *Neuroinformatics* **20**, 25–36 (2022).
26. Moor, M. *et al.* Foundation models for generalist medical artificial intelligence. *Nature* **616**, 259–265 (2023).
27. Hosseini, M., Horbach, S. P. J. M., Holmes, K. & Ross-Hellauer, T. Open Science at the generative AI turn: An exploratory analysis of challenges and opportunities. *Quant. Sci. Stud.* **6**, 22–45 (2025).
28. Liu, J. *et al.* Neural Brain: A neuroscience-inspired framework for embodied agents. *arXiv [cs.RO]* (2025).
29. Caro, J. O. *et al.* BrainLM: A foundation model for brain activity recordings. *bioRxiv* (2023) doi:10.1101/2023.09.12.557460.
30. Luo, X. *et al.* Large language models surpass human experts in predicting neuroscience results. *Nat. Hum. Behav.* **9**, 305–315 (2025).
31. Zeng, Y. *et al.* CellFM: a large-scale foundation model pre-trained on transcriptomics of 100 million human cells. *Nat. Commun.* **16**, 4679 (2025).
32. Talukder, S., Yue, Y. & Gkioxari, G. TOTEM: TOkenized Time Series EMbeddings for general time series

analysis. *arXiv [cs.LG]* (2024).

33. Guo, F. *et al.* Foundation models in bioinformatics. *Natl. Sci. Rev.* **12**, nwaf028 (2025).

34. Schrimpf, M. *et al.* Integrative Benchmarking to Advance Neurally Mechanistic Models of Human Intelligence. *Neuron* **108**, 413–423 (2020).

35. Kell, A. J. E., Yamins, D. L. K., Shook, E. N., Norman-Haignere, S. V. & McDermott, J. H. A Task-Optimized Neural Network Replicates Human Auditory Behavior, Predicts Brain Responses, and Reveals a Cortical Processing Hierarchy. *Neuron* **98**, 630-644.e16 (2018).

36. Mahmood, U., Fu, Z., Ghosh, S., Calhoun, V. & Plis, S. Through the looking glass: Deep interpretable dynamic directed connectivity in resting fMRI. *Neuroimage* **264**, 119737 (2022).

37. Johnson, E. C. *et al.* SciOps: Achieving productivity and reliability in data-intensive research. *arXiv [q-bio.NC]* (2023).

38. Gottweis, J. *et al.* Accelerating scientific discovery with Co-Scientist. *Nature* (2026) doi:10.1038/s41586-026-10644-y.

39. Ghareeb, A. E. *et al.* A multi-agent system for automating scientific discovery. *Nature* (2026) doi:10.1038/s41586-026-10652-y.

40. International Brain Laboratory. Electronic address: churchland@cshl.edu & International Brain Laboratory. An international laboratory for systems and computational neuroscience. *Neuron* **96**, 1213–1218 (2017).

41. Paulk, A. C. *et al.* Large-scale neural recordings with single neuron resolution using Neuropixels probes in human cortex. *Nat. Neurosci.* **25**, 252–263 (2022).

42. Glaser, A. *et al.* Expansion-assisted selective plane illumination microscopy for nanoscale imaging of centimeter-scale tissues. (2025) doi:10.7554/elife.91979.3.

43. Park, J. *et al.* Integrated platform for multiscale molecular imaging and phenotyping of the human brain. *Science* **384**, eadh9979 (2024).

44. Schottdorf, M., Yu, G. & Walker, E. Y. Data science and its future in large neuroscience collaborations. *Neuron* **112**, 3007–3012 (2024).

45. Moore, J. *et al.* OME-Zarr: a cloud-optimized bioimaging file format with international community support. *Histochem. Cell Biol.* **160**, 223–251 (2023).

46. Westner, B. U. *et al.* Cycling on the Freeway: The perilous state of open-source neuroscience software.

*Imaging Neurosci. (Camb.)* **3**, (2025).

47. Katz, D. S., Jensen, E. A. & Barker, M. Understanding and advancing research software grant funding models. *Open Res. Eur.* **5**, 199 (2025).
48. List, J. A. Optimally generate policy-based evidence before scaling. *Nature* **626**, 491–499 (2024).
49. Jalali, M. S. & Hasgul, Z. Potential trade-offs of proposed cuts to the US National Institutes of health. *JAMA Health Forum* **6**, e252228 (2025).
50. Wilkinson, M. D. *et al.* The FAIR Guiding Principles for scientific data management and stewardship. *Sci Data* **3**, 160018 (2016).
51. Carroll, S. R. *et al.* The CARE Principles for Indigenous Data Governance. *CODATA* **19**, 43–43 (2020).
52. Lin, D. *et al.* The TRUST Principles for digital repositories. *Sci. Data* **7**, 144 (2020).
53. Claude Science beta. *Claude* https://claude.com/product/claude-science.
54. Moreau, L. & Missier, P. PROV-DM: The PROV Data Model. (World Wide Web Consortium, 2013).
55. The Computational Democracy Project. https://compdemocracy.org/.
56. Urchs, S. G. W. *et al.* Neurobagel: building an international network for distributed data discovery. *BITSS* (2026) doi:10.31222/osf.io/x64tn_v1.
57. Ai2 Asta. https://asta.allen.ai/.
58. Kosmos: An AI Scientist for Autonomous Discovery. https://edisonscientific.com/news/announcing-kosmos.
59. IGoR. *ARPA-H* https://www.arpa-h.gov/explore-funding/programs/igor.
60. Better Code, Better Science — Better Code, Better Science. https://poldrack.github.io/BetterCodeBetterScience/frontmatter.html.